\documentclass[pra,showpacs,twocolumn,superscriptaddress,amsmath,amssymb]{revtex4-1}
\usepackage{bm}
\usepackage{graphicx}
\usepackage{color}
\usepackage{dcolumn,amsmath}
\usepackage[normalem]{ulem}
\usepackage[version=4]{mhchem}

\begin{document}

\title{Symmetry and the Geometric Phase in Ultracold Hydrogen-Exchange Reactions}

\author{J. F. E. Croft}
\author{J. Hazra}
\author{N. Balakrishnan}
\affiliation{Department of Chemistry, University of Nevada, Las Vegas, Nevada 89154, USA}
\author{B. K. Kendrick}
\affiliation{Theoretical Division (T-1, MS B221), Los Alamos National Laboratory, Los Alamos,
New Mexico 87545, USA}

\begin{abstract}
Quantum reactive scattering calculations are reported for the ultracold
hydrogen-exchange reaction and its non-reactive atom-exchange isotopic
counterparts, proceeding from excited rotational states.
It is shown that while the geometric phase (GP) does not necessarily control the reaction
to all final states one can always find final states where it does.
For the isotopic counterpart reactions these states can be used to make a
measurement of the GP effect by separately measuring the even and odd symmetry
contributions, which experimentally requires nuclear-spin final-state resolution.
This follows from symmetry considerations that make the even and odd
identical-particle exchange symmetry wavefunctions which include the GP locally
equivalent to the opposite symmetry wavefunctions which do not.
This equivalence reflects the important role discrete symmetries play in ultracold chemistry
generally and highlights the key role ultracold reactions can
play in understanding fundamental aspects of chemical reactivity.
\end{abstract}
\maketitle

The hydrogen exchange reaction is referred to as ``the simplest reaction''.
Consisting of only 3 protons and 3 electrons the forces on
the nuclei can be accurately calculated from first-principles quantum mechanics.
Consequently this reaction has been extensively studied which has led to many
advances in our understanding of chemical reactions~\cite{aoiz.banares.ea:h--h2,yang:state-to-state,zare:hydrogen}.

In the ultracold regime chemical reactions can be studied at the single
quantum state level~\cite{ospelkaus.ni.ea:quantum-state,knoop.ferlaino.ea:magnetically,
krems:cold,bell.softley:ultracold,quemener.julienne:ultracold,
balakrishnan:perspective}.
Reactions proceed through a single partial wave and quantum mechanical effects
are magnified. The essence of all chemical reactions is quantum mechanical,
as such the ultracold regime is a window on reactions at their most fundamental level.
A perfect illustration of this is the ultracold reaction between two fermionic KRb molecules,
Ospelkaus {\it et al} showed that the requirement that the total
wavefunction be anti-symmetric with respect to exchange of identical
fermions suppresses the reaction between two KRb molecules in the
same internal state~\cite{ospelkaus.ni.ea:quantum-state}.
Taking advantage of this they were able to turn the reaction
on and off by changing the internal state of one of the molecules.

Studying the ultracold hydrogen exchange reaction
therefore offers the perfect testbed to study fundamental aspects of chemical
reactivity, such as symmetry effects, isotopic substitution and
the GP effect.
The GP effect is purely quantum mechanical in origin,
relating to the phase of the wavefunction encircling a conical intersection (CI)
\cite{longuet-higgins.opik.ea:studies,herzberg.longuet-higgins:intersection,mead.truhlar:on,juanes-marcos.althorpe.ea:theoretical,althorpe:general}.
Being quantum mechanical in origin the GP does not readily manifest itself at
higher collision energies where high quantum numbers lead to classical behaviour
\cite{jankunas.sneha.ea:hunt}.
However in the ultracold regime it has been shown that the GP controls
the \ce{O + OH -> H + O2} reaction~\cite{kendrick.hazra.ea:geometric},
just as the identical particle symmetry does in the KRb reaction.

In quantum mechanics two wavefunctions can interfere either constructively
or destructively, depending on their relative sign (or phase).
A change of sign for either one of the wavefunctions will change destructive
interference to constructive interference or vice versa.
Often such a change of sign is a consequence of symmetry considerations.
This is the case for the hydrogen exchange reaction where the inclusion of the
GP introduces a change of relative sign between the dominant
reaction pathways~\cite{mead:superposition}.
In the ultracold regime phases are quantized, leading to maximally constructive
or destructive interference between reaction pathways.
Furthermore, when two reaction pathways are of similar magnitude they will either
cancel each other out or double up: the reaction \emph{can only be on or off}.
This is exactly the case for the ultracold hydrogen exchange reaction
proceeding from $v=4$ $j=0$, where the sign change associated with the GP
turns the reaction on and off~\cite{kendrick.hazra.ea:geometric*1,kendrick.hazra.ea:geometric*3}.
This is what is meant by the GP controlling ultracold reactions~\cite{kendrick.hazra.ea:geometric}.

In each of these ultracold reactions the on/off character is due to a discrete symmetry.
Discrete symmetries are of fundamental importance in quantum mechanics, but
have no corresponding classical physical meaning.
In this work we examine the ultracold hydrogen exchange reaction, and its non-reactive atom-exchange isotopic
counterparts, proceeding from excited rotational states.
In doing so we will  highlight the important role discrete symmetries play in ultracold
chemistry and under what conditions we should expect them to control ultracold
chemical reactions in general.

\begin{figure*}[t]
\frame{\includegraphics[width=0.8\textwidth]{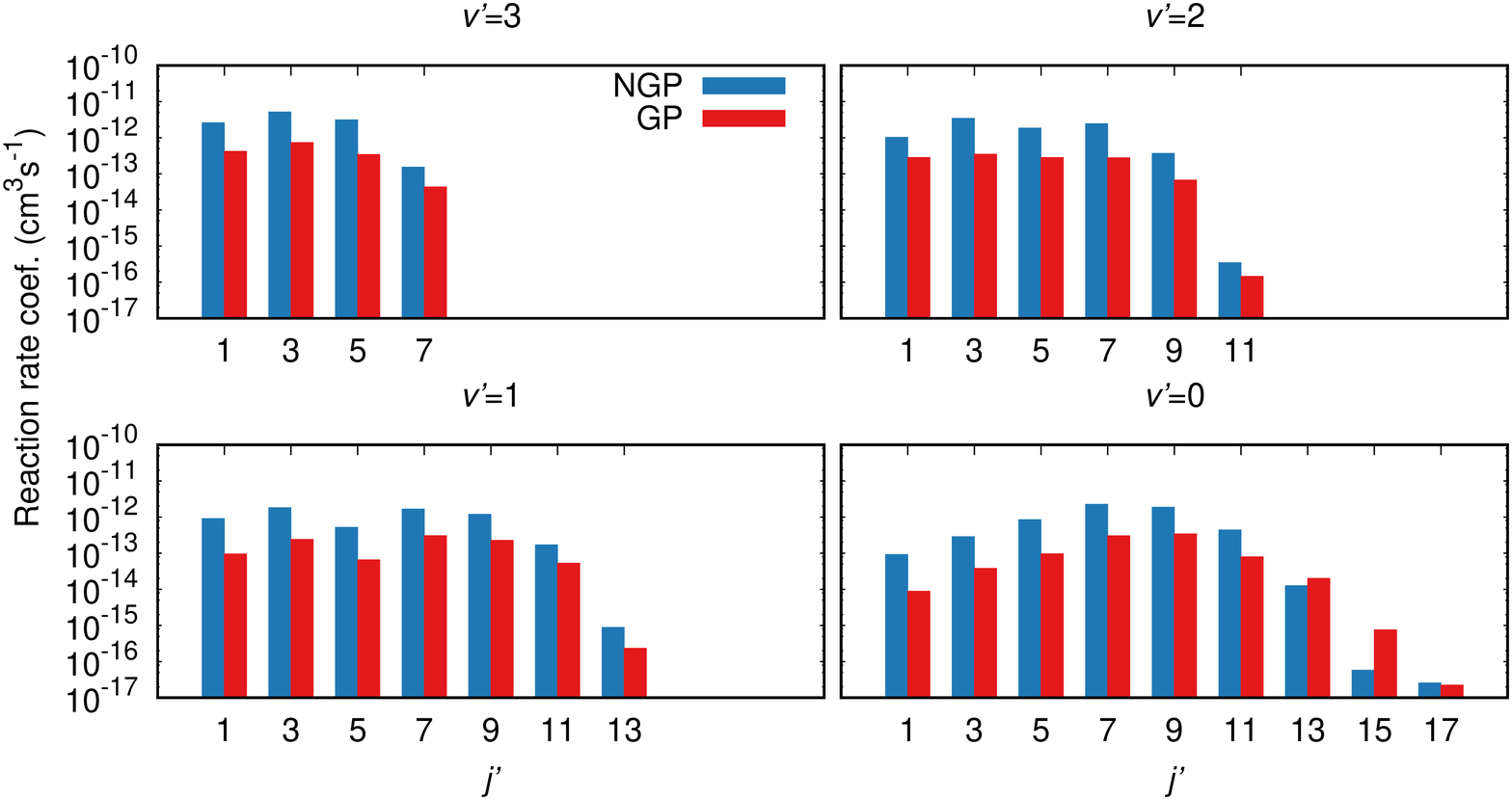}}
\frame{\includegraphics[width=0.8\textwidth]{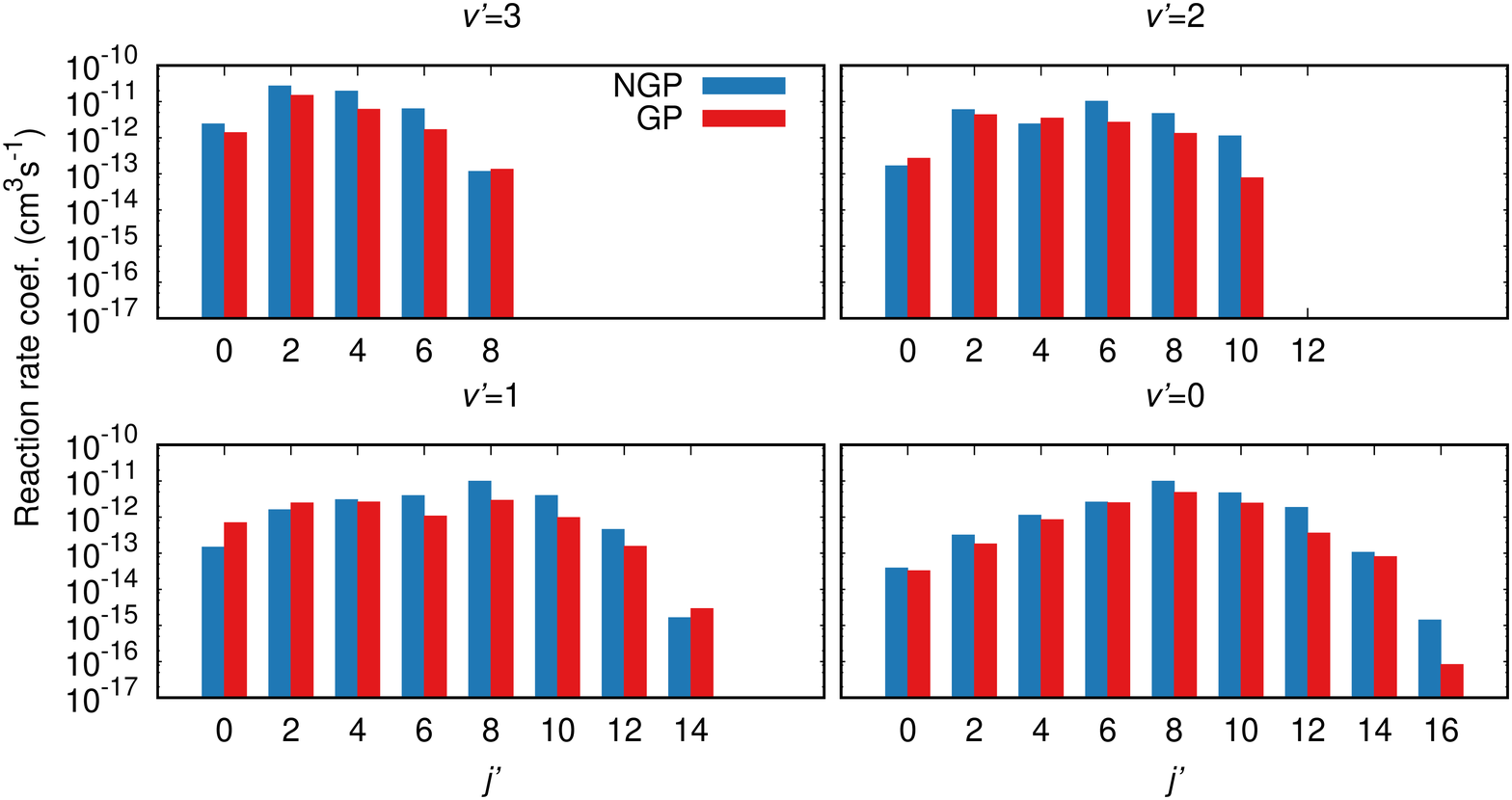}}
\caption{Reaction rate coefficients for the
\ce{H + H2($j=1,2$) -> H + H2($v',j'$)} reaction at 1~$\mu$K.
Results include all values of total angular momentum ($J$) up to and including 4.
The upper and lower panels show reactions proceeding from $j$ =1 and 2 respectively.
The GP and NGP labels denote rates which do and do not include the GP effect
respectively.
}
\label{fig:jresolved_rates_h3}
\end{figure*}

\begin{figure*}[t]
\frame{\includegraphics[width=0.8\textwidth]{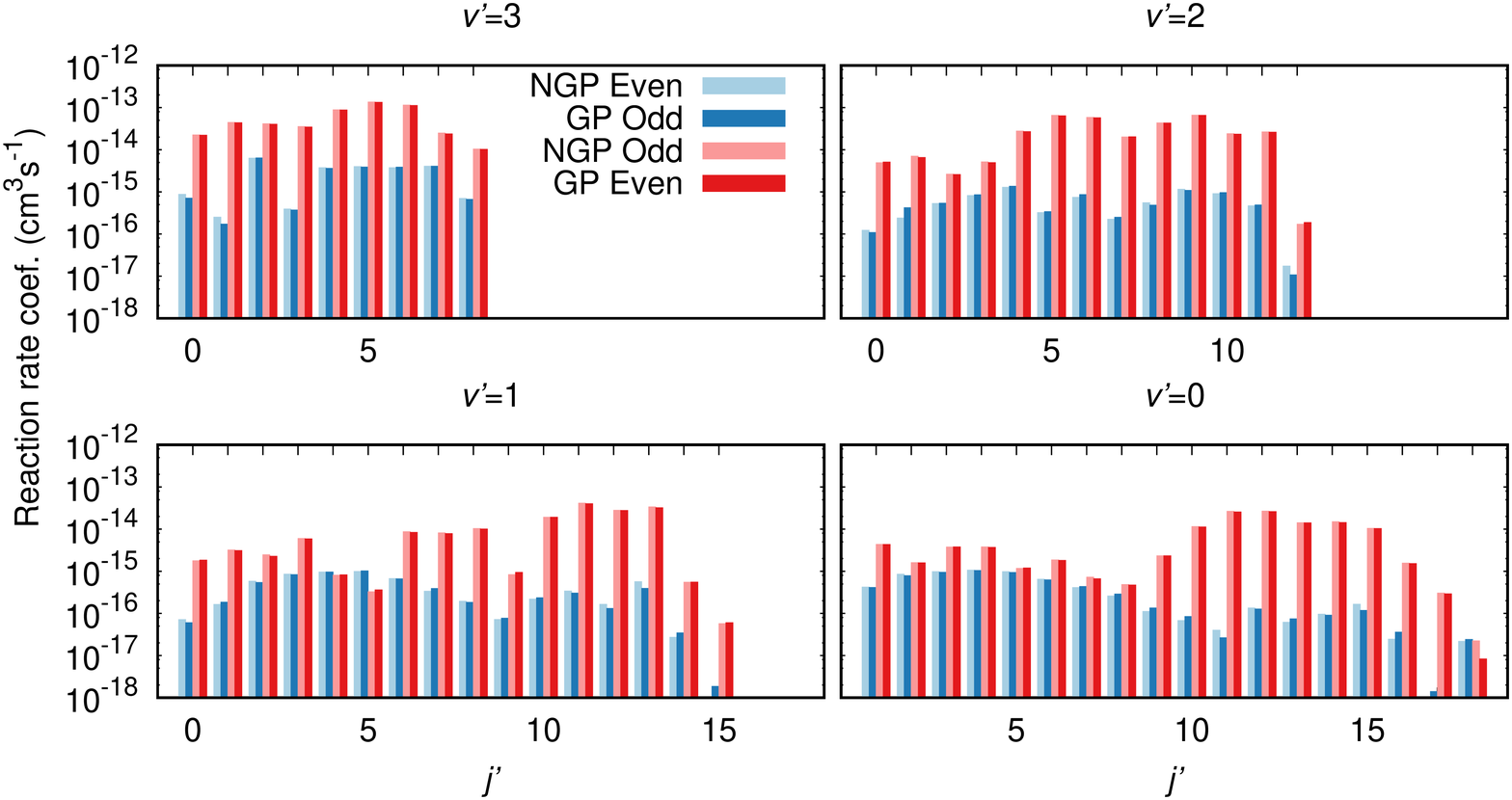}}
\frame{\includegraphics[width=0.8\textwidth]{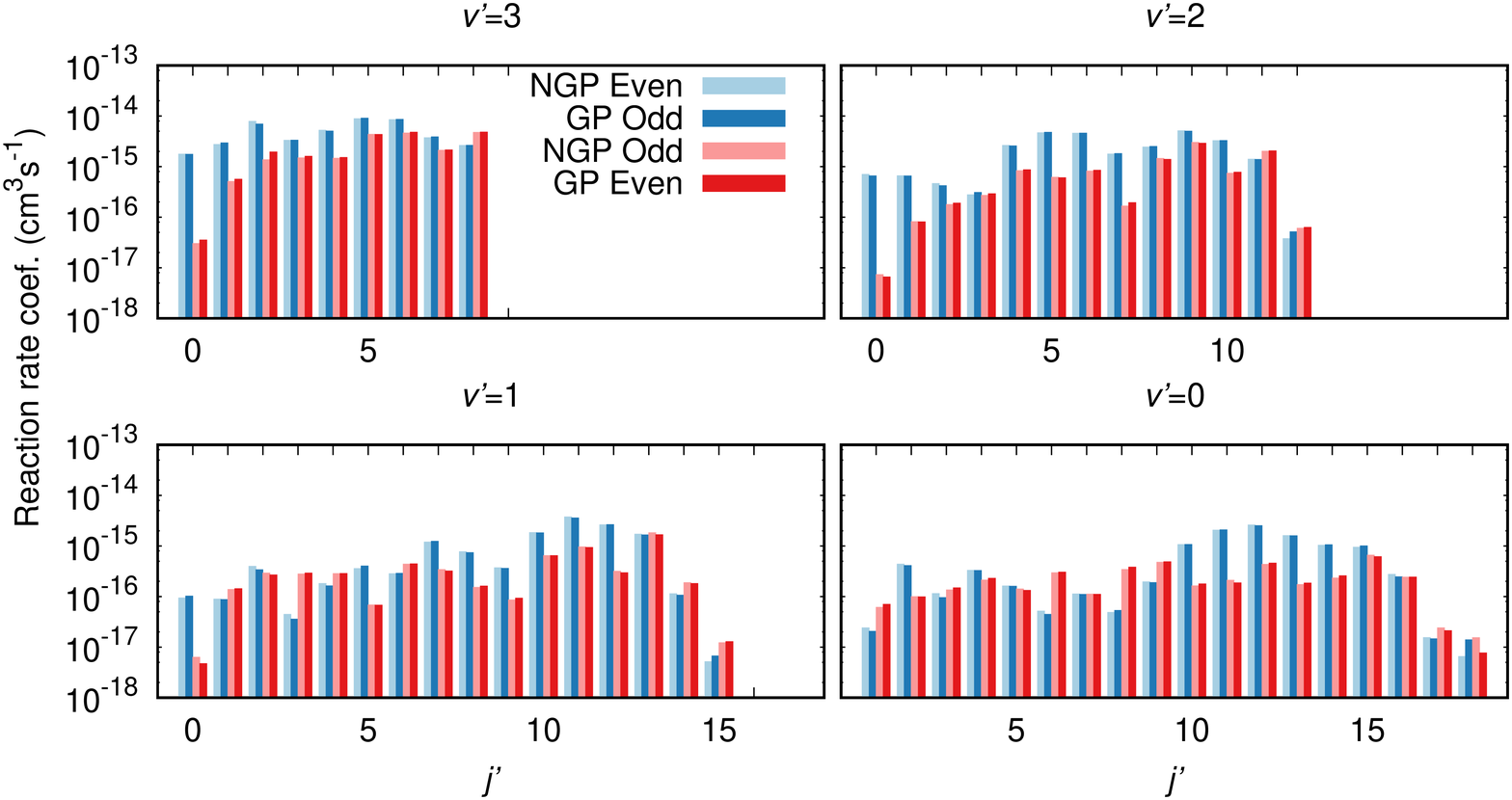}}
\caption{Reaction rate coefficients for the
\ce{D + HD($j=1,2$) -> D + HD($v',j'$)} reaction at 1~$\mu$K.
Results include all values of total angular momentum ($J$) up to and including 4.
The upper and lower panels show reactions proceeding from $j$ =1 and 2 respectively.
}
\label{fig:jresolved_rates_dhd}
\end{figure*}

\begin{figure*}[t]
\frame{\includegraphics[width=0.8\textwidth]{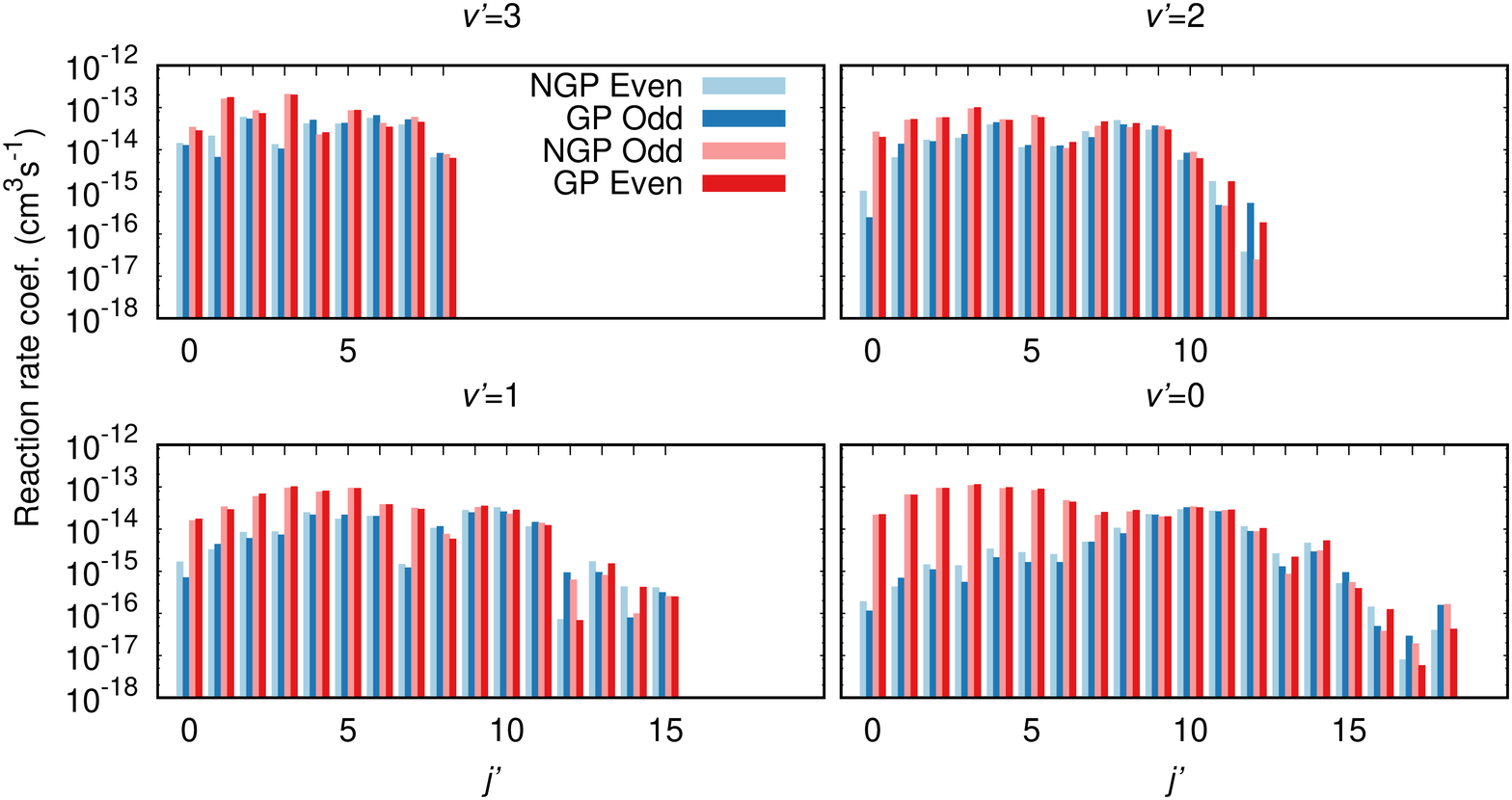}}
\frame{\includegraphics[width=0.8\textwidth]{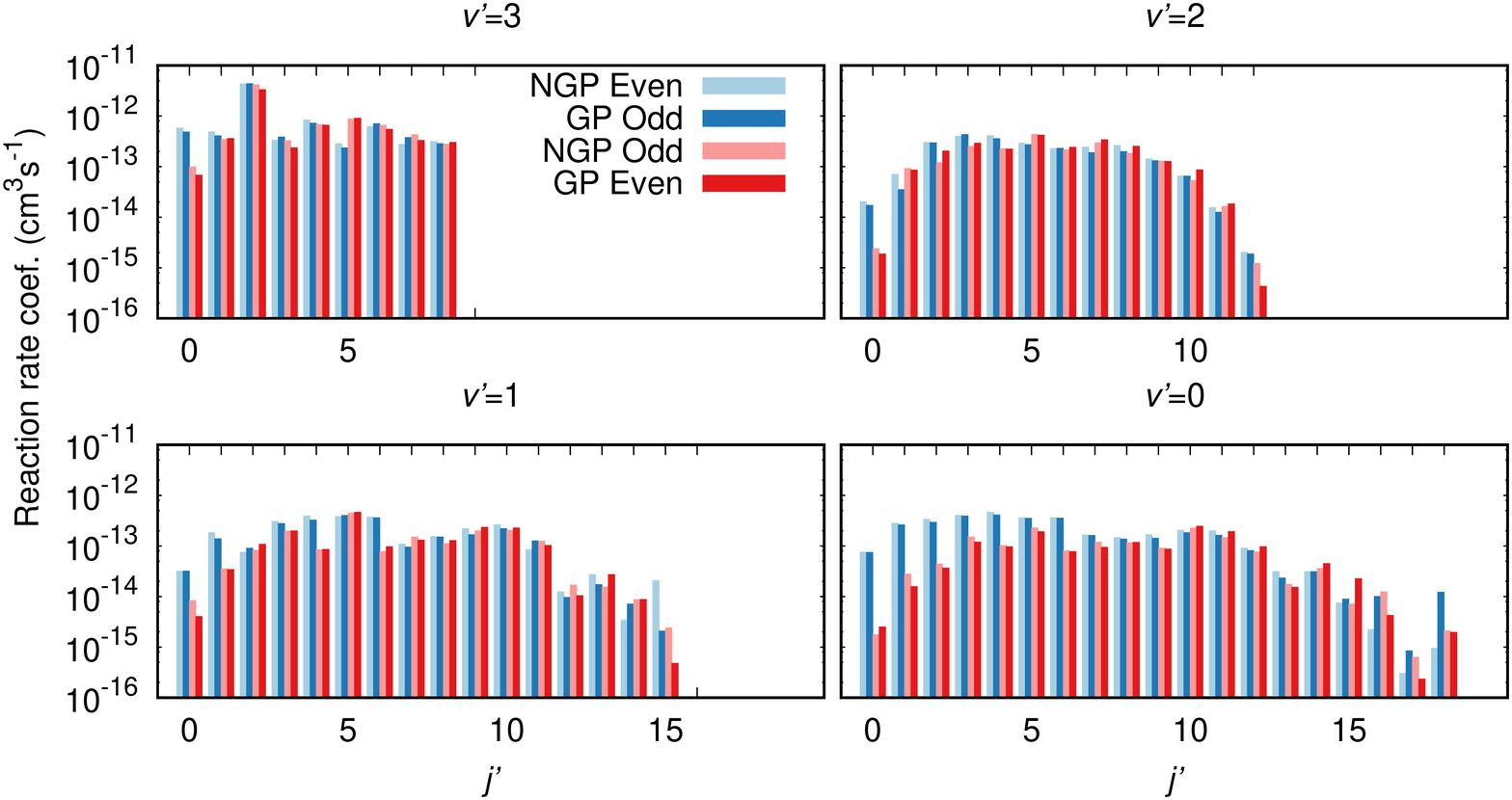}}
\caption{Reaction rate coefficients for the
\ce{H + HD($j=1,2$) -> H + HD($v',j'$)} reaction at 1~$\mu$K.
Results include all values of total angular momentum ($J$) up to and including 4.
The upper and lower panels show reactions proceeding from $j$ =1 and 2 respectively.
}
\label{fig:jresolved_rates_hhd}
\end{figure*}

\begin{figure*}[t]
\frame{\includegraphics[width=0.8\textwidth]{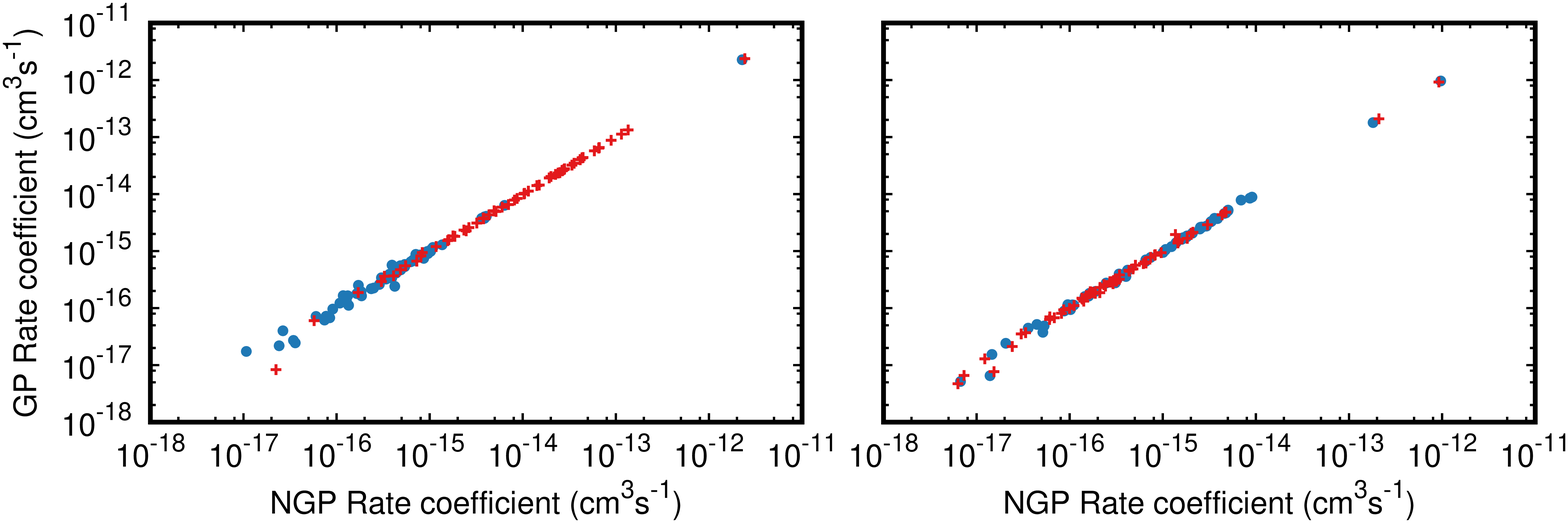}}
\frame{\includegraphics[width=0.8\textwidth]{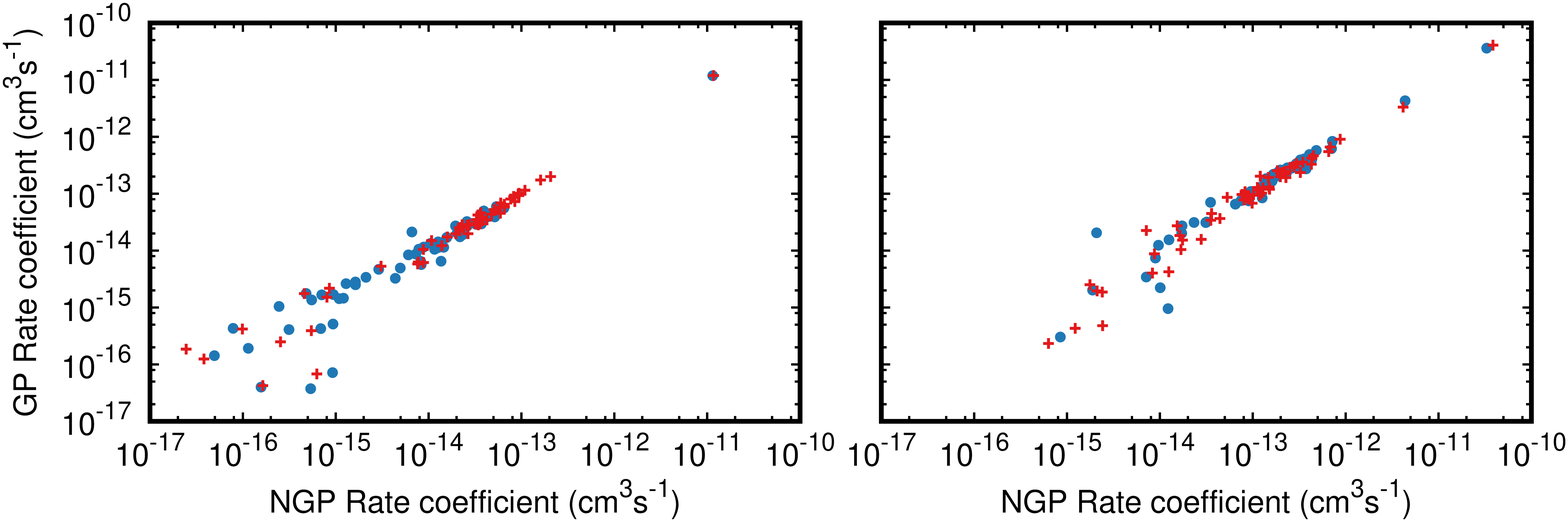}}
\caption{GP rate coefficient vs NGP rate coefficient of the opposite symmetry.
D + HD proceeding from $j=1$ top left,
D + HD proceeding from $j=2$ top right,
H + HD proceeding from $j=1$ bottom left,
and H + HD proceeding from $j=2$ bottom right.
The red crosses correspond to GP even NGP odd while the blue dots correspond to GP odd NGP even.
}
\label{fig:corr}
\end{figure*}

\section{Methods}
We use the atom-diatom scattering formalism as developed by Pack and Parker
\cite{pack.parker:quantum,kendrick.pack.ea:hyperspherical}.
In the short range we use adiabatically-adjusting-principle-axis hyperspherical
coordinates, an approach which ensures that all arrangements are treated fully
equivalently, while in the long range we use Delves hyperspherical coordinates
for each arrangement channel.
Calculations were performed on the BKMP2 potential energy
surface~\cite{boothroyd.keogh.ea:refined} and
the vector potential approach of Mead and Truhler was used to include the GP
effect~\cite{mead.truhlar:on}.
The coupled equations were propagated using the log-derivative method of
Johnson~\cite{johnson:multichannel}.
Results are well converged in the ultracold regime with total angular
momentum up to and including 4 used in all calculations.
This approach has been used extensively in recent years to study
the role of the GP in the ultracold hydrogen exchange reaction and its
isotopic counterparts~\cite{kendrick.hazra.ea:geometric*1,kendrick.hazra.ea:geometric*3,kendrick.hazra.ea:geometric*2,hazra.kendrick.ea:geometric}.

\section{Results}
\subsection{\ce{H + H2 -> H + H2}}
Ultracold reactions proceeding from $v=4$ $j=0$ have been shown to be
controlled by the GP~\cite{kendrick.hazra.ea:geometric*1,kendrick.hazra.ea:geometric*3}.
The excited vibrational state is needed to overcome the reaction barrier while
remaining in the ultracold regime.
As such all reactions in this paper proceed from $v=4$ and the $v$ is omitted from
state-to-state labels for clarity.
As discussed in the introduction the GP controls these reaction due to a number of factors:
the quantization of phase shifts in the ultracold;
the sign change due to the GP;
and the similar magnitude of the two dominant reaction pathways.
In this work we will use the term ``controlled by the GP'' to mean state
resolved rates which change by over an order of magnitude when the GP
is included. Such state-to-state rates offer an excellent way to directly
measure the GP effect in chemical reactions.
We begin by studying the ultracold hydrogen exchange reaction proceeding from
excited rotational states.

Figure~\ref{fig:jresolved_rates_h3} shows $j$-resolved reaction rate coefficients
for \ce{H + H2($j=1,2$) -> H + H2($v',j'$)} calculated at 1~$\mu$K,
in the Wigner threshold regime.
These rates include contributions from both the exchange and non-exchange pathways~\cite{mead:superposition}.
It is seen that for reactions proceeding from $j=1$ the reaction is, just as for $j=0$,
either on or off. The GP and NGP (no geometric phase) rates differ by about an order of magnitude,
with the NGP reaction being on and the GP turning the reaction off.
A couple of final states buck this trend, notably
$v'=0$, $j'=15$ where this is reversed. However these correspond to final
states with rates orders of magnitude smaller than for the dominant final states.
As such the vibrational resolved rates also show this on/off
character.
Reactions proceeding from $j=2$ however show weaker influence of the GP\@.
Here we find that while many channels do not exhibit a strong GP effect many do,
such as $v'=2$, $j'=10$.

H$_2$ exists in either para or ortho form,
for $j=1$ there is no pure rotational quenching whereas for $j=2$
there is pure rotational quenching to $j=0$.
We find that the rate for pure rotational quenching is around 2 orders of magnitude
larger than the rates to inelastic channels and does not exhibit a strong
GP effect.
This is because when there is pure rotational quenching the non-exchange pathway
dominates the reaction and the sign change along the exchange pathway due to
the GP has a small effect.

\subsection{\ce{D + HD -> D + HD} \& \ce{H + HD -> H + HD}}
We now move on to examine the non-reactive atom-exchange isotopic counterparts.
These reactions proceeding from $j=0$ have been studied in the ultracold
regime and shown to exhibit large GP effects in the state-to-state rates~\cite{hazra.kendrick.ea:geometric,kendrick.hazra.ea:geometric*2}.
Figure~\ref{fig:jresolved_rates_dhd} shows $j$-resolved reaction rate coefficients
for \ce{D + HD($j=1,2$) -> D + HD($v',j'$)} at 1~$\mu$K.
Just as for H$_3$ these rates include contributions from both the exchange
and non-exchange pathways.
Here we see that for $j=1$ the GP still controls the reaction,
for even symmetry it turns the reaction on while for odd symmetry it turns
the reaction off.
For reactions proceeding from $j=2$ the trend is mostly reversed and weaker,
however there are many final states where the GP changes the rate by over
an order of magnitude.

Figure~\ref{fig:jresolved_rates_hhd} shows the same data as figure~\ref{fig:jresolved_rates_dhd}
but for the H + HD($j=1,2$) $\rightarrow$ H + HD($v',j'$) reaction.
In this case there is not a consistent trend across all final states.
However there are regions where the GP controls the reaction,
most clearly for $v'=0$ $j' \lesssim 5$ where for $j$=1 (2) the GP turns
on (off) the even symmetry case and turns off (on) the odd symmetry case.
Just as in the H$_3$ case we find that pure rotational quenching dominates
the rate to inelastic channels and does not exhibit a strong GP effect
for either of the non-reactive atom-exchange isotopic counterparts
(here $j=1 \rightarrow$ 0 and $j=2 \rightarrow$ 0 \& 1 are allowed).

It is clear from figures~\ref{fig:jresolved_rates_dhd} and~\ref{fig:jresolved_rates_hhd}
that the GP and exchange symmetry play complementary roles in the non-reactive atom-exchange isotopic
counterparts to the hydrogen exchange reaction.
The GP rates for each symmetry are well approximated by the NGP rate of the
opposite symmetry, this is most clear when the GP controls the reaction
but it is also true generally.
This complementarity follows from the symmetry of the wavefunction
when the GP is included.
The characteristic of the GP is that the wavefunction around the CI is
double valued and exhibits even symmetry on one-side and odd symmetry
on the other side.
The double-valued GP wavefunction can therefore be accurately represented by a
NGP wavefunction (of either even or odd symmetry) but \emph{only locally}
(i.e.\@ only on one side of the CI or the other but not both simultaneously).
A NGP wavefunction of suitable symmetry therefore accurately approximates the
double-valued GP wavefunction in H$_3$ type systems as only one side of
the CI is accessible.
The other side, corresponding to the two transition state pathway, has negligible amplitude
\cite{mead:superposition,kendrick.pack:geometric,kendrick.pack:geometric*1,kendrick:geometric*2,kendrick.pack.ea:hyperspherical,babikov.kendrick.ea:cyclic-n3}.
This equivalence is shown explicitly in figure~\ref{fig:corr} which plots the GP
rate vs the NGP rate of the opposite symmetry and is seen to be particularly
good for the D + HD case.
The points in the top right correspond to the rates for pure rotational quenching
which are orders of magnitude larger than for the other inelastic channels.
These rates will dominate the total rate and do not exhibit
a strong GP effect.

The total state resolved rates are computed by
adding the rates for even and odd exchange symmetry multiplied by the appropriate
nuclear spin statistical factor. This summation reduces the overall
GP effect in the total rates since the rate is on for one of the symmetries
and off for the other.
For the D + HD case since D is a spin 1 boson the even and odd factors are
2/3 and 1/3 respectively.
For the H + HD case since H is a spin 1/2 fermion the even and odd factors
are 1/4 and 3/4 respectively.
The difference between the GP and NGP total rates
is thus primarily due to the nuclear spin weighting factor.

On the other hand with nuclear-spin final-state
resolution the even and odd symmetry GP rates can be measured directly.
The rates shown in figures~\ref{fig:jresolved_rates_dhd} and~\ref{fig:jresolved_rates_hhd}
are then obtained by multiplying by the appropriate nuclear spin weighting factors.
For example the total experimental rate for D + HD is
$k^{\mathrm{exp}}_{\mathrm{tot}} = \frac{2}{3} k^{\mathrm{GP}}_{\mathrm{evn}} + \frac{1}{3} k^{\mathrm{GP}}_{\mathrm{odd}}$.
If instead the experiment measures the rate for a given symmetry, say even, then
$k^{\mathrm{exp}}_{\mathrm{evn}}= \frac{2}{3} k^{\mathrm{GP}}_{\mathrm{evn}}$ and so $k^{\mathrm{GP}}_{\mathrm{evn}}=\frac{3}{2} k^{\mathrm{exp}}_{\mathrm{evn}}$,
while for the odd case $k^{\mathrm{exp}}_{\mathrm{odd}}=\frac{1}{3} k^{\mathrm{GP}}_{\mathrm{odd}}$ and so $k^{\mathrm{GP}}_{\mathrm{odd}}= \frac{3}{1} k^{\mathrm{exp}}_{\mathrm{odd}}$.
Due to the symmetry correspondence between the GP and NGP rates,
the experimentally measured rates are well approximated by the NGP rates of opposite symmetry:
$k^{\mathrm{NGP}}_{\mathrm{odd}}=\frac{3}{2}\,k^{\mathrm{exp}}_{\mathrm{evn}}$ and $k^{\mathrm{NGP}}_{\mathrm{evn}}=\frac{3}{1}\,k^{\mathrm{exp}}_{\mathrm{odd}}$.
In contrast, a calculation which ignores the GP entirely would predict:
 $k^{\mathrm{NGP}}_{\mathrm{odd}}=\frac{3}{2}\,k^{\mathrm{exp}}_{\mathrm{odd}}$ and $k^{\mathrm{NGP}}_{\mathrm{evn}}=\frac{3}{1}\,k^{\mathrm{exp}}_{\mathrm{evn}}$.
While the examples given here are in the Wigner threshold regime this
is quite general, in fact for the non-reactive atom-exchange isotopic counterparts
to the hydrogen exchange reaction \emph{any state resolved rate at any energy
exhibiting a significant difference between the even and odd exchange symmetry
is exhibiting a strong GP effect}.

\section{Conclusions}
We have examined the hydrogen exchange reaction proceeding from initial
states $v=4$ $j=1$ \& 2, finding that the GP plays a crucial role in the reaction.
For reactions proceeding from $j=1$ the GP plays an important role in all
final states whereas for the $j=2$ the effect is reduced but there are still
many final states where the GP has a strong effect.

For the non-reactive atom-exchange isotopic counterparts to the hydrogen exchange reaction,
just as in the H$_3$ case, we find that there are always final
states which exhibit a strong GP effect.
Due to symmetry, for these reactions, state-to-state rates including the GP
are well approximated by NGP rates of the opposite identical-particle
exchange symmetry.
This symmetry effect can be used to make a measurement of the GP effect.
Experimentally this amounts to finding a final state with a large
difference between even and odd symmetry, which requires nuclear-spin
final-state resolution.

The importance of the GP and identical-particle symmetry shown here reflects the
importance of discrete symmetries in ultracold chemical reactions generally,
where their effect is magnified.
Discrete symmetries have been shown to play an key role in a diverse range
of ultracold reactions: \ce{KRb + KRb -> K2 + Rb2}
\cite{ospelkaus.ni.ea:quantum-state}, \ce{O + OH -> O2 + H}~\cite{kendrick.hazra.ea:geometric},
and the hydrogen exchange reaction~\cite{kendrick.hazra.ea:geometric*1}.
Discrete symmetries are present in all quantum systems exhibiting
reflection symmetry and this on/off character is expected to be
ubiquitous across ultracold chemistry.
This highlights the important role ultracold reactions can play in understanding
fundamental chemical processes more generally.

\section{Acknowledgments}
We acknowledge support from the  US Army Research Office, MURI grant No.~W911NF-12-1-0476
(N.B.), the US National Science Foundation, grant No.~PHY-1505557 (N.B.).
BKK acknowledges that part of this
work was done under the auspices of the US Department of Energy,
Project No. 20170221ER of the Laboratory Directed Research and Development
Program at Los Alamos National Laboratory.
Los Alamos National Laboratory is operated by Los Alamos National Security,
LLC, for the National Security Administration of the US Department of Energy
under contract DE-AC52-06NA25396.

\bibliography{../../all}

\begin{thebibliography}{29}%
\makeatletter
\providecommand \@ifxundefined [1]{%
 \@ifx{#1\undefined}
}%
\providecommand \@ifnum [1]{%
 \ifnum #1\expandafter \@firstoftwo
 \else \expandafter \@secondoftwo
 \fi
}%
\providecommand \@ifx [1]{%
 \ifx #1\expandafter \@firstoftwo
 \else \expandafter \@secondoftwo
 \fi
}%
\providecommand \natexlab [1]{#1}%
\providecommand \enquote  [1]{``#1''}%
\providecommand \bibnamefont  [1]{#1}%
\providecommand \bibfnamefont [1]{#1}%
\providecommand \citenamefont [1]{#1}%
\providecommand \href@noop [0]{\@secondoftwo}%
\providecommand \href [0]{\begingroup \@sanitize@url \@href}%
\providecommand \@href[1]{\@@startlink{#1}\@@href}%
\providecommand \@@href[1]{\endgroup#1\@@endlink}%
\providecommand \@sanitize@url [0]{\catcode `\\12\catcode `\$12\catcode
  `\&12\catcode `\#12\catcode `\^12\catcode `\_12\catcode `\%12\relax}%
\providecommand \@@startlink[1]{}%
\providecommand \@@endlink[0]{}%
\providecommand \url  [0]{\begingroup\@sanitize@url \@url }%
\providecommand \@url [1]{\endgroup\@href {#1}{\urlprefix }}%
\providecommand \urlprefix  [0]{URL }%
\providecommand \Eprint [0]{\href }%
\providecommand \doibase [0]{http://dx.doi.org/}%
\providecommand \selectlanguage [0]{\@gobble}%
\providecommand \bibinfo  [0]{\@secondoftwo}%
\providecommand \bibfield  [0]{\@secondoftwo}%
\providecommand \translation [1]{[#1]}%
\providecommand \BibitemOpen [0]{}%
\providecommand \bibitemStop [0]{}%
\providecommand \bibitemNoStop [0]{.\EOS\space}%
\providecommand \EOS [0]{\spacefactor3000\relax}%
\providecommand \BibitemShut  [1]{\csname bibitem#1\endcsname}%
\let\auto@bib@innerbib\@empty
\bibitem [{\citenamefont {Aoiz}, \citenamefont {Ba{\~n}ares},\ and\
  \citenamefont {Herrero}(2005)}]{aoiz.banares.ea:h--h2}%
  \BibitemOpen
  \bibfield  {author} {\bibinfo {author} {\bibfnamefont {F.}~\bibnamefont
  {Aoiz}}, \bibinfo {author} {\bibfnamefont {L.}~\bibnamefont {Ba{\~n}ares}}, \
  and\ \bibinfo {author} {\bibfnamefont {V.}~\bibnamefont {Herrero}},\
  }\href@noop {} {\bibfield  {journal} {\bibinfo  {journal} {Int. Rev. Phys.
  Chem.}\ }\textbf {\bibinfo {volume} {24}},\ \bibinfo {pages} {119} (\bibinfo
  {year} {2005})}\BibitemShut {NoStop}%
\bibitem [{\citenamefont {Yang}(2007)}]{yang:state-to-state}%
  \BibitemOpen
  \bibfield  {author} {\bibinfo {author} {\bibfnamefont {X.}~\bibnamefont
  {Yang}},\ }\href@noop {} {\bibfield  {journal} {\bibinfo  {journal} {Annu.
  Rev. Phys. Chem.}\ }\textbf {\bibinfo {volume} {58}},\ \bibinfo {pages} {433}
  (\bibinfo {year} {2007})}\BibitemShut {NoStop}%
\bibitem [{\citenamefont {Zare}(2013)}]{zare:hydrogen}%
  \BibitemOpen
  \bibfield  {author} {\bibinfo {author} {\bibfnamefont {R.~N.}\ \bibnamefont
  {Zare}},\ }\href@noop {} {\bibfield  {journal} {\bibinfo  {journal} {Annu.
  Rev. Phys. Chem.}\ }\textbf {\bibinfo {volume} {64}},\ \bibinfo {pages} {1}
  (\bibinfo {year} {2013})}\BibitemShut {NoStop}%
\bibitem [{\citenamefont {Ospelkaus}\ \emph {et~al.}(2010)\citenamefont
  {Ospelkaus}, \citenamefont {Ni}, \citenamefont {Wang}, \citenamefont
  {de~Miranda}, \citenamefont {Neyenhuis}, \citenamefont {Qu{\'e}m{\'e}ner},
  \citenamefont {Julienne}, \citenamefont {Bohn}, \citenamefont {Jin},\ and\
  \citenamefont {Ye}}]{ospelkaus.ni.ea:quantum-state}%
  \BibitemOpen
  \bibfield  {author} {\bibinfo {author} {\bibfnamefont {S.}~\bibnamefont
  {Ospelkaus}}, \bibinfo {author} {\bibfnamefont {K.-K.}\ \bibnamefont {Ni}},
  \bibinfo {author} {\bibfnamefont {D.}~\bibnamefont {Wang}}, \bibinfo {author}
  {\bibfnamefont {M.~H.~G.}\ \bibnamefont {de~Miranda}}, \bibinfo {author}
  {\bibfnamefont {B.}~\bibnamefont {Neyenhuis}}, \bibinfo {author}
  {\bibfnamefont {G.}~\bibnamefont {Qu{\'e}m{\'e}ner}}, \bibinfo {author}
  {\bibfnamefont {P.~S.}\ \bibnamefont {Julienne}}, \bibinfo {author}
  {\bibfnamefont {J.~L.}\ \bibnamefont {Bohn}}, \bibinfo {author}
  {\bibfnamefont {D.~S.}\ \bibnamefont {Jin}}, \ and\ \bibinfo {author}
  {\bibfnamefont {J.}~\bibnamefont {Ye}},\ }\href {\doibase
  10.1126/science.1184121} {\bibfield  {journal} {\bibinfo  {journal}
  {Science}\ }\textbf {\bibinfo {volume} {327}},\ \bibinfo {pages} {853}
  (\bibinfo {year} {2010})}\BibitemShut {NoStop}%
\bibitem [{\citenamefont {Knoop}\ \emph {et~al.}(2010)\citenamefont {Knoop},
  \citenamefont {Ferlaino}, \citenamefont {Berninger}, \citenamefont {Mark},
  \citenamefont {N\"agerl}, \citenamefont {Grimm}, \citenamefont {D'Incao},\
  and\ \citenamefont {Esry}}]{knoop.ferlaino.ea:magnetically}%
  \BibitemOpen
  \bibfield  {author} {\bibinfo {author} {\bibfnamefont {S.}~\bibnamefont
  {Knoop}}, \bibinfo {author} {\bibfnamefont {F.}~\bibnamefont {Ferlaino}},
  \bibinfo {author} {\bibfnamefont {M.}~\bibnamefont {Berninger}}, \bibinfo
  {author} {\bibfnamefont {M.}~\bibnamefont {Mark}}, \bibinfo {author}
  {\bibfnamefont {H.-C.}\ \bibnamefont {N\"agerl}}, \bibinfo {author}
  {\bibfnamefont {R.}~\bibnamefont {Grimm}}, \bibinfo {author} {\bibfnamefont
  {J.~P.}\ \bibnamefont {D'Incao}}, \ and\ \bibinfo {author} {\bibfnamefont
  {B.~D.}\ \bibnamefont {Esry}},\ }\href {\doibase
  10.1103/PhysRevLett.104.053201} {\bibfield  {journal} {\bibinfo  {journal}
  {Phys. Rev. Lett.}\ }\textbf {\bibinfo {volume} {104}},\ \bibinfo {pages}
  {053201} (\bibinfo {year} {2010})}\BibitemShut {NoStop}%
\bibitem [{\citenamefont {Krems}(2008)}]{krems:cold}%
  \BibitemOpen
  \bibfield  {author} {\bibinfo {author} {\bibfnamefont {R.~V.}\ \bibnamefont
  {Krems}},\ }\href@noop {} {\bibfield  {journal} {\bibinfo  {journal} {Phys.
  Chem. Chem. Phys.}\ }\textbf {\bibinfo {volume} {10}},\ \bibinfo {pages}
  {4079} (\bibinfo {year} {2008})}\BibitemShut {NoStop}%
\bibitem [{\citenamefont {Bell}\ and\ \citenamefont
  {Softley}(2009)}]{bell.softley:ultracold}%
  \BibitemOpen
  \bibfield  {author} {\bibinfo {author} {\bibfnamefont {M.~T.}\ \bibnamefont
  {Bell}}\ and\ \bibinfo {author} {\bibfnamefont {T.~P.}\ \bibnamefont
  {Softley}},\ }\href@noop {} {\bibfield  {journal} {\bibinfo  {journal} {Mol.
  Phys.}\ }\textbf {\bibinfo {volume} {107}},\ \bibinfo {pages} {99} (\bibinfo
  {year} {2009})}\BibitemShut {NoStop}%
\bibitem [{\citenamefont {Qu\'{e}m\'{e}ner}\ and\ \citenamefont
  {Julienne}(2012)}]{quemener.julienne:ultracold}%
  \BibitemOpen
  \bibfield  {author} {\bibinfo {author} {\bibfnamefont {G.}~\bibnamefont
  {Qu\'{e}m\'{e}ner}}\ and\ \bibinfo {author} {\bibfnamefont {P.~S.}\
  \bibnamefont {Julienne}},\ }\href {\doibase 10.1021/cr300092g} {\bibfield
  {journal} {\bibinfo  {journal} {Chem. Rev.}\ }\textbf {\bibinfo {volume}
  {112}},\ \bibinfo {pages} {4949} (\bibinfo {year} {2012})}\BibitemShut
  {NoStop}%
\bibitem [{\citenamefont {Balakrishnan}(2016)}]{balakrishnan:perspective}%
  \BibitemOpen
  \bibfield  {author} {\bibinfo {author} {\bibfnamefont {N.}~\bibnamefont
  {Balakrishnan}},\ }\href {\doibase 10.1063/1.4964096} {\bibfield  {journal}
  {\bibinfo  {journal} {J. Chem. Phys.}\ }\textbf {\bibinfo {volume} {145}},\
  \bibinfo {pages} {150901} (\bibinfo {year} {2016})}\BibitemShut {NoStop}%
\bibitem [{\citenamefont {Longuet-Higgins}\ \emph {et~al.}(1958)\citenamefont
  {Longuet-Higgins}, \citenamefont {Opik}, \citenamefont {Pryce},\ and\
  \citenamefont {Sack}}]{longuet-higgins.opik.ea:studies}%
  \BibitemOpen
  \bibfield  {author} {\bibinfo {author} {\bibfnamefont {H.}~\bibnamefont
  {Longuet-Higgins}}, \bibinfo {author} {\bibfnamefont {U.}~\bibnamefont
  {Opik}}, \bibinfo {author} {\bibfnamefont {M.}~\bibnamefont {Pryce}}, \ and\
  \bibinfo {author} {\bibfnamefont {R.}~\bibnamefont {Sack}},\ }in\ \href@noop
  {} {\emph {\bibinfo {booktitle} {Proc. Roy. Soc. Lond. A}}},\ Vol.\ \bibinfo
  {volume} {244}\ (\bibinfo {organization} {The Royal Society},\ \bibinfo
  {year} {1958})\ pp.\ \bibinfo {pages} {1--16}\BibitemShut {NoStop}%
\bibitem [{\citenamefont {Herzberg}\ and\ \citenamefont
  {Longuet-Higgins}(1963)}]{herzberg.longuet-higgins:intersection}%
  \BibitemOpen
  \bibfield  {author} {\bibinfo {author} {\bibfnamefont {G.}~\bibnamefont
  {Herzberg}}\ and\ \bibinfo {author} {\bibfnamefont {H.~C.}\ \bibnamefont
  {Longuet-Higgins}},\ }\href@noop {} {\bibfield  {journal} {\bibinfo
  {journal} {Discuss. Faraday Soc.}\ }\textbf {\bibinfo {volume} {35}},\
  \bibinfo {pages} {77} (\bibinfo {year} {1963})}\BibitemShut {NoStop}%
\bibitem [{\citenamefont {Mead}\ and\ \citenamefont
  {Truhlar}(1979)}]{mead.truhlar:on}%
  \BibitemOpen
  \bibfield  {author} {\bibinfo {author} {\bibfnamefont {C.~A.}\ \bibnamefont
  {Mead}}\ and\ \bibinfo {author} {\bibfnamefont {D.~G.}\ \bibnamefont
  {Truhlar}},\ }\href@noop {} {\bibfield  {journal} {\bibinfo  {journal} {J.
  Chem. Phys.}\ }\textbf {\bibinfo {volume} {70}},\ \bibinfo {pages} {2284}
  (\bibinfo {year} {1979})}\BibitemShut {NoStop}%
\bibitem [{\citenamefont {Juanes-Marcos}, \citenamefont {Althorpe},\ and\
  \citenamefont {Wrede}(2005)}]{juanes-marcos.althorpe.ea:theoretical}%
  \BibitemOpen
  \bibfield  {author} {\bibinfo {author} {\bibfnamefont {J.~C.}\ \bibnamefont
  {Juanes-Marcos}}, \bibinfo {author} {\bibfnamefont {S.~C.}\ \bibnamefont
  {Althorpe}}, \ and\ \bibinfo {author} {\bibfnamefont {E.}~\bibnamefont
  {Wrede}},\ }\href {\doibase 10.1126/science.1114890} {\bibfield  {journal}
  {\bibinfo  {journal} {Science}\ }\textbf {\bibinfo {volume} {309}},\ \bibinfo
  {pages} {1227} (\bibinfo {year} {2005})}\BibitemShut {NoStop}%
\bibitem [{\citenamefont {Althorpe}(2006)}]{althorpe:general}%
  \BibitemOpen
  \bibfield  {author} {\bibinfo {author} {\bibfnamefont {S.~C.}\ \bibnamefont
  {Althorpe}},\ }\href@noop {} {\bibfield  {journal} {\bibinfo  {journal} {J.
  Chem. Phys.}\ }\textbf {\bibinfo {volume} {124}},\ \bibinfo {pages} {084105}
  (\bibinfo {year} {2006})}\BibitemShut {NoStop}%
\bibitem [{\citenamefont {Jankunas}\ \emph {et~al.}(2013)\citenamefont
  {Jankunas}, \citenamefont {Sneha}, \citenamefont {Zare}, \citenamefont
  {Bouakline},\ and\ \citenamefont {Althorpe}}]{jankunas.sneha.ea:hunt}%
  \BibitemOpen
  \bibfield  {author} {\bibinfo {author} {\bibfnamefont {J.}~\bibnamefont
  {Jankunas}}, \bibinfo {author} {\bibfnamefont {M.}~\bibnamefont {Sneha}},
  \bibinfo {author} {\bibfnamefont {R.~N.}\ \bibnamefont {Zare}}, \bibinfo
  {author} {\bibfnamefont {F.}~\bibnamefont {Bouakline}}, \ and\ \bibinfo
  {author} {\bibfnamefont {S.~C.}\ \bibnamefont {Althorpe}},\ }\href {\doibase
  10.1063/1.4821601} {\bibfield  {journal} {\bibinfo  {journal} {J. Chem.
  Phys.}\ }\textbf {\bibinfo {volume} {139}},\ \bibinfo {pages} {144316}
  (\bibinfo {year} {2013})}\BibitemShut {NoStop}%
\bibitem [{\citenamefont {Kendrick}, \citenamefont {Hazra},\ and\ \citenamefont
  {Balakrishnan}(2015{\natexlab{a}})}]{kendrick.hazra.ea:geometric}%
  \BibitemOpen
  \bibfield  {author} {\bibinfo {author} {\bibfnamefont {B.}~\bibnamefont
  {Kendrick}}, \bibinfo {author} {\bibfnamefont {J.}~\bibnamefont {Hazra}}, \
  and\ \bibinfo {author} {\bibfnamefont {N.}~\bibnamefont {Balakrishnan}},\
  }\href@noop {} {\bibfield  {journal} {\bibinfo  {journal} {Nat. Commun.}\
  }\textbf {\bibinfo {volume} {6}},\ \bibinfo {pages} {7918} (\bibinfo {year}
  {2015}{\natexlab{a}})}\BibitemShut {NoStop}%
\bibitem [{\citenamefont {Mead}(1980)}]{mead:superposition}%
  \BibitemOpen
  \bibfield  {author} {\bibinfo {author} {\bibfnamefont {C.~A.}\ \bibnamefont
  {Mead}},\ }\href {\doibase 10.1063/1.439600} {\bibfield  {journal} {\bibinfo
  {journal} {J. Chem. Phys.}\ }\textbf {\bibinfo {volume} {72}},\ \bibinfo
  {pages} {3839} (\bibinfo {year} {1980})}\BibitemShut {NoStop}%
\bibitem [{\citenamefont {Kendrick}, \citenamefont {Hazra},\ and\ \citenamefont
  {Balakrishnan}(2015{\natexlab{b}})}]{kendrick.hazra.ea:geometric*1}%
  \BibitemOpen
  \bibfield  {author} {\bibinfo {author} {\bibfnamefont {B.~K.}\ \bibnamefont
  {Kendrick}}, \bibinfo {author} {\bibfnamefont {J.}~\bibnamefont {Hazra}}, \
  and\ \bibinfo {author} {\bibfnamefont {N.}~\bibnamefont {Balakrishnan}},\
  }\href {\doibase 10.1103/PhysRevLett.115.153201} {\bibfield  {journal}
  {\bibinfo  {journal} {Phys. Rev. Lett.}\ }\textbf {\bibinfo {volume} {115}},\
  \bibinfo {pages} {153201} (\bibinfo {year} {2015}{\natexlab{b}})}\BibitemShut
  {NoStop}%
\bibitem [{\citenamefont {Kendrick}, \citenamefont {Hazra},\ and\ \citenamefont
  {Balakrishnan}(2016{\natexlab{a}})}]{kendrick.hazra.ea:geometric*3}%
  \BibitemOpen
  \bibfield  {author} {\bibinfo {author} {\bibfnamefont {B.~K.}\ \bibnamefont
  {Kendrick}}, \bibinfo {author} {\bibfnamefont {J.}~\bibnamefont {Hazra}}, \
  and\ \bibinfo {author} {\bibfnamefont {N.}~\bibnamefont {Balakrishnan}},\
  }\href {\doibase 10.1063/1.4966037} {\bibfield  {journal} {\bibinfo
  {journal} {J. Chem. Phys.}\ }\textbf {\bibinfo {volume} {145}},\ \bibinfo
  {pages} {164303} (\bibinfo {year} {2016}{\natexlab{a}})}\BibitemShut
  {NoStop}%
\bibitem [{\citenamefont {Pack}\ and\ \citenamefont
  {Parker}(1987)}]{pack.parker:quantum}%
  \BibitemOpen
  \bibfield  {author} {\bibinfo {author} {\bibfnamefont {R.~T.}\ \bibnamefont
  {Pack}}\ and\ \bibinfo {author} {\bibfnamefont {G.~A.}\ \bibnamefont
  {Parker}},\ }\href {\doibase http://dx.doi.org/10.1063/1.452944} {\bibfield
  {journal} {\bibinfo  {journal} {J. Chem. Phys.}\ }\textbf {\bibinfo {volume}
  {87}},\ \bibinfo {pages} {3888} (\bibinfo {year} {1987})}\BibitemShut
  {NoStop}%
\bibitem [{\citenamefont {Kendrick}\ \emph {et~al.}(1999)\citenamefont
  {Kendrick}, \citenamefont {Pack}, \citenamefont {Walker},\ and\ \citenamefont
  {Hayes}}]{kendrick.pack.ea:hyperspherical}%
  \BibitemOpen
  \bibfield  {author} {\bibinfo {author} {\bibfnamefont {B.~K.}\ \bibnamefont
  {Kendrick}}, \bibinfo {author} {\bibfnamefont {R.~T.}\ \bibnamefont {Pack}},
  \bibinfo {author} {\bibfnamefont {R.~B.}\ \bibnamefont {Walker}}, \ and\
  \bibinfo {author} {\bibfnamefont {E.~F.}\ \bibnamefont {Hayes}},\ }\href@noop
  {} {\bibfield  {journal} {\bibinfo  {journal} {J. Chem. Phys.}\ }\textbf
  {\bibinfo {volume} {110}},\ \bibinfo {pages} {6673} (\bibinfo {year}
  {1999})}\BibitemShut {NoStop}%
\bibitem [{\citenamefont {Boothroyd}\ \emph {et~al.}(1996)\citenamefont
  {Boothroyd}, \citenamefont {Keogh}, \citenamefont {Martin},\ and\
  \citenamefont {Peterson}}]{boothroyd.keogh.ea:refined}%
  \BibitemOpen
  \bibfield  {author} {\bibinfo {author} {\bibfnamefont {A.~I.}\ \bibnamefont
  {Boothroyd}}, \bibinfo {author} {\bibfnamefont {W.~J.}\ \bibnamefont
  {Keogh}}, \bibinfo {author} {\bibfnamefont {P.~G.}\ \bibnamefont {Martin}}, \
  and\ \bibinfo {author} {\bibfnamefont {M.~R.}\ \bibnamefont {Peterson}},\
  }\href {\doibase 10.1063/1.471430} {\bibfield  {journal} {\bibinfo  {journal}
  {J. Chem. Phys.}\ }\textbf {\bibinfo {volume} {104}},\ \bibinfo {pages}
  {7139} (\bibinfo {year} {1996})}\BibitemShut {NoStop}%
\bibitem [{\citenamefont {Johnson}(1973)}]{johnson:multichannel}%
  \BibitemOpen
  \bibfield  {author} {\bibinfo {author} {\bibfnamefont {B.~R.}\ \bibnamefont
  {Johnson}},\ }\href@noop {} {\bibfield  {journal} {\bibinfo  {journal} {J.
  Comput. Phys.}\ }\textbf {\bibinfo {volume} {13}},\ \bibinfo {pages} {445}
  (\bibinfo {year} {1973})}\BibitemShut {NoStop}%
\bibitem [{\citenamefont {Kendrick}, \citenamefont {Hazra},\ and\ \citenamefont
  {Balakrishnan}(2016{\natexlab{b}})}]{kendrick.hazra.ea:geometric*2}%
  \BibitemOpen
  \bibfield  {author} {\bibinfo {author} {\bibfnamefont {B.~K.}\ \bibnamefont
  {Kendrick}}, \bibinfo {author} {\bibfnamefont {J.}~\bibnamefont {Hazra}}, \
  and\ \bibinfo {author} {\bibfnamefont {N.}~\bibnamefont {Balakrishnan}},\
  }\href@noop {} {\bibfield  {journal} {\bibinfo  {journal} {New J. Phys.}\
  }\textbf {\bibinfo {volume} {18}},\ \bibinfo {pages} {123020} (\bibinfo
  {year} {2016}{\natexlab{b}})}\BibitemShut {NoStop}%
\bibitem [{\citenamefont {Hazra}, \citenamefont {Kendrick},\ and\ \citenamefont
  {Balakrishnan}(2016)}]{hazra.kendrick.ea:geometric}%
  \BibitemOpen
  \bibfield  {author} {\bibinfo {author} {\bibfnamefont {J.}~\bibnamefont
  {Hazra}}, \bibinfo {author} {\bibfnamefont {B.~K.}\ \bibnamefont {Kendrick}},
  \ and\ \bibinfo {author} {\bibfnamefont {N.}~\bibnamefont {Balakrishnan}},\
  }\href@noop {} {\bibfield  {journal} {\bibinfo  {journal} {J. Phys. B}\
  }\textbf {\bibinfo {volume} {49}},\ \bibinfo {pages} {194004} (\bibinfo
  {year} {2016})}\BibitemShut {NoStop}%
\bibitem [{\citenamefont {Kendrick}\ and\ \citenamefont
  {Pack}(1996)}]{kendrick.pack:geometric}%
  \BibitemOpen
  \bibfield  {author} {\bibinfo {author} {\bibfnamefont {B.}~\bibnamefont
  {Kendrick}}\ and\ \bibinfo {author} {\bibfnamefont {R.~T.}\ \bibnamefont
  {Pack}},\ }\href {\doibase 10.1063/1.471460} {\bibfield  {journal} {\bibinfo
  {journal} {J. Chem. Phys.}\ }\textbf {\bibinfo {volume} {104}},\ \bibinfo
  {pages} {7475} (\bibinfo {year} {1996})}\BibitemShut {NoStop}%
\bibitem [{\citenamefont {Kendrick}\ and\ \citenamefont
  {Pack}(1997)}]{kendrick.pack:geometric*1}%
  \BibitemOpen
  \bibfield  {author} {\bibinfo {author} {\bibfnamefont {B.}~\bibnamefont
  {Kendrick}}\ and\ \bibinfo {author} {\bibfnamefont {R.~T.}\ \bibnamefont
  {Pack}},\ }\href {\doibase 10.1063/1.473449} {\bibfield  {journal} {\bibinfo
  {journal} {J. Chem. Phys.}\ }\textbf {\bibinfo {volume} {106}},\ \bibinfo
  {pages} {3519} (\bibinfo {year} {1997})}\BibitemShut {NoStop}%
\bibitem [{\citenamefont {Kendrick}(2003)}]{kendrick:geometric*2}%
  \BibitemOpen
  \bibfield  {author} {\bibinfo {author} {\bibfnamefont {B.~K.}\ \bibnamefont
  {Kendrick}},\ }\href@noop {} {\bibfield  {journal} {\bibinfo  {journal} {J.
  Phys. Chem. A}\ }\textbf {\bibinfo {volume} {107}},\ \bibinfo {pages} {6739}
  (\bibinfo {year} {2003})}\BibitemShut {NoStop}%
\bibitem [{\citenamefont {Babikov}\ \emph {et~al.}(2005)\citenamefont
  {Babikov}, \citenamefont {Kendrick}, \citenamefont {Zhang},\ and\
  \citenamefont {Morokuma}}]{babikov.kendrick.ea:cyclic-n3}%
  \BibitemOpen
  \bibfield  {author} {\bibinfo {author} {\bibfnamefont {D.}~\bibnamefont
  {Babikov}}, \bibinfo {author} {\bibfnamefont {B.~K.}\ \bibnamefont
  {Kendrick}}, \bibinfo {author} {\bibfnamefont {P.}~\bibnamefont {Zhang}}, \
  and\ \bibinfo {author} {\bibfnamefont {K.}~\bibnamefont {Morokuma}},\ }\href
  {\doibase 10.1063/1.1824905} {\bibfield  {journal} {\bibinfo  {journal} {J.
  Chem. Phys.}\ }\textbf {\bibinfo {volume} {122}},\ \bibinfo {pages} {044315}
  (\bibinfo {year} {2005})}\BibitemShut {NoStop}%
\end{thebibliography}%

\end{document}